\documentclass[12pt]{article}
\newcommand{\sect}[1]{\section{#1}\setcounter{equation}{0}}

\begin{document}
\bigskip
\hspace*{\fill}
\vbox{\baselineskip12pt \hbox{UCSBTH-97-21}\hbox{hep-th/9710158}}
\bigskip\bigskip\bigskip

\centerline{\Large \bf Singularities in wavy strings}
\bigskip\bigskip

\centerline{\large Simon F. Ross\footnote{\tt
sross@cosmic.physics.ucsb.edu}}
\medskip
\centerline{Physics Department}
\centerline{University of California}
\centerline{Santa Barbara, CA 93106}
\bigskip\bigskip

\begin{abstract}
Extremal six-dimensional black string solutions with some non-trivial
momentum distribution along the wave are considered. These solutions
were recently shown to contain a singularity at the would-be position
of the event horizon.  In the black string geometry, all curvature
invariants are finite at the horizon. It is shown that if the effects
of infalling matter are included, there are curvature invariants which
diverge there. This implies that quantum corrections will be important
at the would-be horizon. The effect of this singularity on test
strings is also considered, and it is shown that it leads to a
divergent excitation of the string. The quantum corrections will
therefore be important for test objects.
\end{abstract}
 
\newpage

\sect{Introduction}

In the last two years, sensational progress in the understanding of
the thermodynamics of black holes has been achieved. A statistical
understanding of the Bekenstein-Hawking entropy for a large number of
black holes has been obtained. That is, it is possible to find states
in string theory at weak coupling whose strong coupling limit
corresponds to a black hole, and the number of states that correspond
to a particular black hole is given by the exponential of the black
hole's entropy (for reviews, see \cite{hor:stringc,mald:D-ent}). The
first successful counting was obtained for extreme five-dimensional
black holes \cite{strom:D-ent}. One can think of these as extreme
black strings in six dimensions, with the direction along the string
compacitified. One way to generalize this solution is by adding
traveling waves moving along the string, i.e., by making the momentum
distribution along the string inhomogeneous
\cite{gar:cswave,gar:tw}. In \cite{hor:tw1,hor:tw2}, it was shown that
the statistical understanding of the Bekenstein-Hawking entropy can be
extended to these black strings with traveling waves.  This is an
impressive success for the string picture, as the entropy depends on
arbitrary functions describing the wave.

This success seems somewhat mysterious, in light of the later
observation \cite{hor:twsing,kaloper:twsing} that the would-be event
horizon of the black strings with traveling waves is
singular. However, this singularity appears to be a fairly mild
one. The total tidal distortion of an infalling body remains
finite. In a thermal ensemble of black strings with a given total
momentum, the typical string will have only small inhomogeneities in
the momentum distribution. For such a typical black string, the total
tidal distortion only differs from that occasioned by a string without
traveling waves by a factor of order one \cite{mar:seff}. Furthermore,
the Ricci tensor and all the curvature invariants remain finite at the
singularity, which was used in \cite{hor:twsing} to argue that
$\alpha'$ corrections to this solution would be suppressed.

This paper presents a further investigation of the nature of this
singularity. I will first show that if we consider the effects of
infalling matter on the geometry, invariants which diverge as the
singularity at the would-be horizon is approached can be
constructed. Thus, it seems likely that if one consistently
incorporates the back-reaction of infalling matter, the $\alpha'$
corrections and perturbative quantum gravity corrections become
important. This is similar to the analysis of certain near-extreme
black holes in \cite{hor:prop}, but in the present case, one can
construct a divergent invariant as soon as there is any amount of
infalling matter. This result applies for any type of matter.

The experience of \cite{hor:prop} indicates that the existence of
these divergent invariants will not fundamentally alter the motion of
test particles. One might therefore wonder if the singularity is still
mild in a certain sense, even if quantum corrections are
important. However, in string theory, the appropriate test objects are
not particles following geodesics, but rather first-quantized test
strings \cite{hor:wave}. In section \ref{app}, I show that from the
point of view of a family of infalling observers, the metric looks
approximately like that of a plane wave. Thus, the propagation of test
strings can be approximated by the propagation in a plane wave. The
expectation value of the mass squared operator $\langle M_>^2 \rangle$
diverges on propagating through the singularity at the
horizon. Although I only discuss the six-dimensional black string with
longitudinal waves, similar conclusions should hold for the string
with internal waves, and for the five-dimensional black
string. Therefore, at least for strings, the classical analysis of the
motion of test objects will break down as we approach the would-be
horizon, and quantum corrections will have an important effect.

The singularity at the would-be horizon is, from these two points of
view, far from mild. This makes it seem even more mysterious that one
can successfully reproduce the entropy of these objects using string
theory. Even though the geometry near the horizon will receive large
quantum corrections, it gives the correct density of states. On the
other hand, the existence of large quantum corrections near the event
horizon should have important consequences for the information loss
problem in the context of these solutions, particularly as these
corrections have an important effect on infalling observers. These
solutions thus provide a new perspective on both the relation between
geometry and entropy, and the information loss problem. The most
important caveat is that this discussion only covers the extreme
solutions, and it is not clear that this analysis can be extended to
non-extreme configurations.

In the next section, I will review the six-dimensional black string
solution, following \cite{hor:twsing} closely. The curvature near the
would-be horizon has a form which is reminiscent of a plane wave. In
section \ref{backr}, I show that combining the stress-energy of
infalling matter with the curvature, one can obtain invariants which
diverge at the singularity. In section \ref{app}, I show that the
metric can be approximated by a plane wave from the point of view of a
suitable family of observers. On propagating through such a plane
wave, the expectation value of the mass squared for a test string
diverges.  Section \ref{concl} contains some speculations on the
implications of these results.

\sect{Black strings with traveling waves}
\label{review}

The solutions considered here are solutions of the low-energy
effective theory obtained from type IIB string theory. The solutions
are expressed in the ten dimensional Einstein frame. Since only the
metric, dilaton, and RR three-form $H$ are non-trivial, they are
solutions of the effective action
\begin{equation} \label{action}
S = \frac{1}{16 \pi G} \int d^{10}x \sqrt{-g} \left[ R - \frac{1}{2}
(\nabla \phi)^2 - \frac{1}{12} e^\phi H^2 \right].
\end{equation}
A class of six-dimensional extremal black string solutions of
(\ref{action}) was studied in \cite{hor:twsing}. For the sake of
simplicity, I will only consider solutions with a longitudinal wave
and no internal wave. This allows us to reduce the solution to the
six-dimensional metric
\begin{equation} \label{twmet1}
ds^2 = - \left( 1 - \frac{r_0^2}{r^2} \right) du dv+ \frac{p(u)}{r^2}
du^2 +  \left( 1 - \frac{r_0^2}{r^2} \right)^{-2} dr^2 + r^2 d\Omega_3.
\end{equation}
Here $v,u=t\pm z$, where $z$ is a coordinate on an $S^1$ with period
$L$, so $p(u)$ is a periodic function of $u$. For this solution,
$\phi=0$, so the Einstein and string frames are identical. 

At the horizon $r=r_0$, this coordinate system breaks down, as $t \to
\infty$ at the horizon, implying $u,v \to \infty$. A coordinate system
in which the metric is $C^0$ at the horizon was found in
\cite{hor:tw1}. This form of the metric is written in terms of a new
function $\sigma(u)$, defined by
\begin{equation} \label{sigma}
\sigma^2(u) + \dot{\sigma}(u) = \frac{p(u)}{r_0^4},
\end{equation}
which implies $\sigma(u)$ is a periodic function with the same period
$L$ as $p(u)$. We also define $R = r/r_0$,
\begin{equation} \label{Gu}
G(u) = e^{\int_0^u \sigma du},
\end{equation}
and
\begin{equation} \label{W}
W = G \left( 1 - \frac{1}{R^2} \right)^{1/2}.
\end{equation}
The necessary coordinate transformations are
\begin{equation} \label{newcoorda}
U = - \int_u^{+\infty} \frac{du}{G^2}, 
\end{equation}
\begin{equation} \label{newcoordb}
q = -\frac{r_0}{2W^2} - 3 r_0\int_0^U \sigma dU,
\end{equation}
\begin{equation} \label{newcoordc}
V = v - \frac{r_0^2 \sigma}{R^2 -1} - 2r_0^2 \int_0^u \sigma^2 du + 3
r_0^2 \int_0^U \sigma^2 W^2 dU.
\end{equation}
The metric (\ref{twmet1}) can be written in terms of these coordinates
as
\begin{eqnarray} \label{twmet2}
ds^2 &=& -W^2 dU dV + r_0^2 \sigma^2 W^4 (R^2-1)(4R^2-3) dU^2
\nonumber \\ &&+ \left[ 2 r_0 \sigma W^4 R^2 (R^2-1)(2R^2+1) + 6 r_0W^2
\int_0^U \sigma^2 W^4 dU \right] dq dU  \\ &&+ W^4 R^6 dq^2 +
R^2 r_0^2 d\Omega_3.  \nonumber
\end{eqnarray}
In this coordinate system, the future event horizon lies at $U=0$, and
all the other coordinates are well-behaved at the event horizon. The
region outside the horizon corresponds to $U<0$. The metric is
independent of $V$; that is, $\partial/\partial V$ is a null Killing
vector. The horizon area is $A=2\pi^2r_0^4\int_0^L \sigma du$. 

To write down the curvature components, \cite{hor:twsing} used a null
sechsbein\footnote{This is the same notation as used in
\cite{hor:twsing}, even though the labels $5$ and $6$ appear to be
reversed relative to (2.10) of \cite{hor:twsing}. This is because they
are actually writing the $(e^\mu)_a$ rather than $(e_\mu)_a$.}
\begin{eqnarray} \label{tet}
(e_1)_a &=& R r_0 \partial_a \theta, \quad (e_2)_a = R r_0 \sin \theta
\partial_a \phi, \quad (e_3)_a = R r_0 \sin \theta \sin \phi \partial_a
\psi, \nonumber \\  
(e_4)_a &=& W^2 R^3 \partial_a q, \quad (e_6)_a = W^2 \partial_a U, \\
(e_5)_a &=& -\frac{1}{2} \partial_a V + \frac{1}{2} \sigma^2 W^2
(R^2-1)(4R^2-3) \partial_a U \nonumber \\ &&+ \left[  r_0 \sigma W^2
R^2(R^2-1)(2R^2+1) + 3 r_0 \int_0^U \sigma^2 W^4 dU \right] \partial_a q. 
\nonumber 
\end{eqnarray}
This is a natural choice of basis, given the metric
(\ref{twmet2}). The angles $\theta, \phi, \psi$ are coordinates on the
three-sphere, and the metric in this basis is $ds^2 = (e_1)^2 +(e_2)^2
+(e_3)^2 +(e_4)^2 +2(e_5)(e_6)$. Some of the curvature components
diverge at the horizon. The divergent terms are
\begin{equation} \label{curva}
R_{1515} = R_{2525} = R_{3535} = \frac{\dot{\sigma}}{R^2-1} +
\mbox{finite part}, 
\end{equation}
and
\begin{equation} \label{curvb}
R_{4545} = -\frac{3\dot{\sigma}}{R^2-1} + \mbox{finite part}, 
\end{equation}
where an overdot denotes $\partial/\partial u$. The other non-zero
curvature components $\sim 1/r_0^2$ or $\sim \sigma/r_0$ near the
horizon. Note that the Ricci tensor will be finite; that is, the
divergent contribution to the curvature comes entirely from the Weyl
tensor. This divergence in the curvature can be thought of
as arising because $\sigma$ is periodic in $u$, so it oscillates an
infinite number of times as the horizon is approached.

If one writes a gravitational plane wave in a null basis, the non-zero
curvature components are just $R_{i5i5}$ (no sum on $i$), and the
Ricci tensor vanishes. Hence, the divergent part of the curvature here
is reminiscent of a gravitational plane wave. It was also shown in
\cite{hor:twsing} that near $U=0$, the curvature $\sim 1/U$. We will
see in section \ref{app} that the metric (\ref{twmet2}) can indeed be
approximated by a plane wave near $U=0$.

\sect{Effects of infalling matter on the geometry}
\label{backr}

In \cite{hor:twsing}, it was argued that the absence of divergent
curvature invariants at the would-be horizon in the black string
metric might indicate that corrections to the classical geometry were
suppressed. In \cite{hor:prop}, it was realized that in metrics with
large null curvatures, the back-reaction of infalling matter could
make an important difference in such considerations. One can construct
large curvature invariants by including the matter stress tensor, even
if the stress tensor remains small. In the six-dimensional black
string metric, the Ricci tensor in the basis (\ref{tet}) is
\begin{equation} \label{ricci}
R_{11}= R_{22} = R_{33} = - R_{44}= -R_{56} = \frac{2}{r_0^2 R^6},
\qquad R_{55} = O(\dot{\sigma}, \sigma).
\end{equation}
Note that in particular $R_{66} = 0$. Since the Ricci tensor doesn't
diverge, we can't find large curvature invariants by considering
$T_{\mu\nu} T^{\mu\nu}$, as was done in \cite{hor:prop}. However,
\begin{equation} \label{rcont}
R^{\mu\nu} R_{\mu 5 \nu 5} = \frac{12 \dot{\sigma}}{r_0^2 R^6 (R^2-1)}
+ \ldots,
\end{equation}
so $T^{\rho \sigma} R^{\mu\nu} R_{\mu \rho \nu \sigma}$ will diverge
at the horizon if the stress tensor of the matter $T_{\rho \sigma}$
contains an ingoing part, that is, if $T_{66} \neq 0$. Adding any
non-zero amount of ingoing matter gives rise to a divergent
invariant. The presence of this divergence is also independent of the
type of matter we use, and of the size of the inhomogeneities on the
black string. This argument extends immediately to the
five-dimensional black string, and to the black string with internal
waves, as in all these cases the analogue of (\ref{rcont}) diverges.

In \cite{hor:prop}, it was found that the presence of such large
modifications of the invariants did not necessarily imply that the
motion of infalling particles in the exact metric would deviate
significantly from that in the original metric. That is, the existence
of these large invariants doesn't invalidate the arguments that the
total tidal distortion of infalling bodies remains
finite. Essentially, this is because the infalling bodies already feel
the large curvature; their tidal distortion is finite simply because
the double integral of the curvature over proper time is
finite. Unfortunately, the complicated nature of the black string
metric precludes any analysis along the lines of \cite{hor:prop} of
the motion of infalling matter in the exact classical solution.

One might object that the divergent object constructed above is not,
technically, a curvature invariant. However, even though it's too
difficult to construct an exact classical solution including the
curvature due to the infalling matter, I expect that it would contain
divergent curvature invariants arising from similar contractions
between the curvature due to infalling matter and the black hole's
curvature.

The presence of large invariants presumably implies that string
$\alpha'$ corrections and perturbative quantum corrections are
becoming important. Since one obtains large invariants as soon as any
infalling matter is added to the solution, such corrections must in
some sense already be important in the original solution, since
quantum fluctuations in the matter fields should be able to produce
the necessary ingoing flux. However, the infalling observers
experience small integrated tidal distortions if the inhomogeneities
on the black string are small enough. One might then argue that the
quantum corrections will have little effect on them.

\sect{Effects on test strings}
\label{app}

In this section, we consider the behavior of test strings in the
singular black string solution. A test string propagating through the
would-be horizon will receive a divergent excitation. Therefore, the
quantum corrections to the metric near the horizon must have an
important effect on these test strings. To simplify the analysis, I
show that the part of the spacetime (\ref{twmet2}) traversed
by small infalling observers can be approximated by a plane wave
metric. The approximation of the metric near the horizon was also
discussed in \cite{mar:seff}.

We want to find a simpler metric which describes the geometry seen by
a suitable family of infalling geodesics. There are not enough
conserved quantities to allow us to determine the geodesics
explicitly. However, $q$ is a good coordinate at the horizon, so we
can choose a family of observers with $q$ approximately constant near
the horizon. For such a family of observers, we can neglect the
dependence on $q$ in the metric. We also wish to neglect the $dU^2$
and $dq dU$ terms in the metric. These terms do not have any effect on
the divergent part of the curvature; indeed, the only part of the
curvature they contribute to is the sub-leading part of the $R_{i5i5}$
terms. We might therefore argue for neglecting them on the grounds
that they have no relevance for the physics we are interested
in. Another weak argument for neglecting the $dq dU$ term is that it
vanishes to leading order, and since $q$ is approximately constant,
the sub-leading part is less important than the $dU dV$ term. I will
give an argument for neglecting the $dU^2$ term later in this
section. The resulting approximate metric will be
\begin{equation} \label{appmet1}
ds^2 \approx -W^2 dU dV + W^4 R^6 dq^2 + R^2 r_0^2 d\Omega_3,
\end{equation}
where $W$ and $R$ are now just functions of $U$. Note that in the
metric (\ref{appmet1}), $p_q = W^4 R^6 \dot{q}$ is a constant of
motion, so there are geodesics of this approximate metric with $q$
constant. 

If we take $r_0\gg1$, we can also approximate the three-sphere metric
$r_0^2 d\Omega_3$ by a flat metric. I will use $x_1, x_2, x_3$ to
denote these flat directions. We should also take $r_0 \gg \sigma$ to
ensure that the finite part of the curvature is negligible. Let us
also define $\tilde{U}$ such that $d\tilde{U}= W^2dU$. The metric is
then
\begin{equation} \label{appmet2}
ds^2 \approx - d\tilde{U} dV + W^4 R^6 dq^2 + R^2 dx_i dx^i.
\end{equation}
This is a plane wave metric. To bring it into the form used in
\cite{hor:wave}, use a change of coordinates discussed in
\cite{gib:wave}, 
\begin{equation} \label{ct1}
\tilde{V} = V + \frac{1}{2}(W^4 R^6)' q^2 + \frac{1}{2} (R^2)' x_i
x^i,  
\end{equation}
\begin{equation} \label{ct2}
X_i = R x_i, \qquad X_4 = W^2 R^3 q,
\end{equation}
where a prime denotes $\partial/\partial \tilde{U}$. 
In terms of these coordinates, the metric (\ref{appmet2}) is
\begin{equation} \label{wavemet}
ds^2 \approx -d\tilde{U} d\tilde{V} + dX_\mu dX^\mu + \left[ \frac{R''}{R}
X_i^2 + \frac{(W^2 R^3)''}{W^2 R^3} X_4^2 \right] d\tilde{U}^2,
\end{equation}
where $\mu$ runs over $1,\ldots, 4$, and $i$ runs over $1,\ldots,
3$. If we had kept the $dU^2$ term from (\ref{twmet1}), we could now
give an argument for neglecting it, since it vanishes at the horizon,
and is therefore negligible compared to the $d\tilde{U}^2$ term in
(\ref{wavemet}).

The geodesics in this metric have a conserved momentum $P=
d{\tilde{U}}/d\tau$ associated with the Killing vector
$\partial/\partial V$, where $\tau$ is the proper time along the
geodesic. Using $d\tilde{U} = W^2 dU$ and (\ref{newcoorda}), we can
write
\begin{equation} \label{pv}
P = \frac{d\tilde{U}}{d\tau} = W^2 \frac{d{U}}{d\tau}  =
\frac{W^2}{G^2} \frac{du}{d\tau} = \left(1-
\frac{1}{R^2}\right)  \frac{du}{d\tau} = p_v,
\end{equation}
where $p_v$ is the momentum associated with the Killing
vector $\partial/\partial v$ of the original form of the metric
(\ref{twmet1}). For geodesics which start from infinity with small
initial velocity, $|P|=|p_v| \approx 1$. Hence $\tilde{U}$ is
approximately equal to the proper time along such geodesics. Note,
however, that I have not shown that there are geodesics which both
start from infinity with small velocity and cross the horizon with $q$
approximately constant. I won't make any particular assumption about
the size of $P$ in the subsequent analysis. 

The metric is approximately a plane wave metric. We should now
consider the form of the coefficient of $d\tilde{U}^2$. Using
$d\tilde{U} = W^2 dU$ and (\ref{newcoorda}),
\begin{equation} \label{rdd}
\frac{R''}{R} = \frac{1}{R}\frac{\partial}{\partial \tilde{U}} \left(
\frac{\partial}{\partial \tilde{U}} R \right) = \frac{G^2}{R W^2}
\frac{\partial}{\partial u} \left( \frac{G^2}{W^2}
\frac{\partial}{\partial u} R\right). 
\end{equation}
From (\ref{W}), it follows that
\begin{equation} \label{rd}
\dot{R} = \frac{WR^3}{G^2} \dot{W} - \frac{W^2 R^3}{G^3} \dot{G},
\end{equation}
and from (\ref{newcoordb}) it follows that
\begin{equation} \label{wd}
\dot{W} = \frac{3 \sigma W^3}{G^2},
\end{equation}
while $\dot{G} = \sigma G$ (here, as before, an overdot denotes
$\partial/\partial u$). Substituting (\ref{rd}) and (\ref{wd}) into
(\ref{rdd}), we find
\begin{equation} \label{div1}
\frac{R''}{R} = \frac{G^2}{R W^2}
\frac{\partial}{\partial u} \left( - \sigma R^3 + \frac{3 W^2 \sigma
R^3}{G^2} \right) = - \frac{\dot{\sigma}}{R^2-1} + \mbox{finite
terms}. 
\end{equation}
Similarly, 
\begin{equation} \label{wrd}
\frac{\partial}{\partial u} (W^2 R^3) = \frac{3 \sigma W^4 R^3}{G^2} +
\frac{6 W^6 \sigma R^5}{G^4}, 
\end{equation}
and hence
\begin{equation} \label{div2}
\frac{(W^2R^3)''}{W^2R^3} = \frac{G^2}{R^3 W^4}
\frac{\partial}{\partial u} \left[ \frac{G^2}{W^2}
\frac{\partial}{\partial u} (W^2 R^3) \right] = \frac{3
\dot{\sigma}}{R^2-1} + \mbox{finite terms}. 
\end{equation}
Note that no approximations were used in obtaining
(\ref{div1},\ref{div2}). We see that the singular terms now appear
directly in the metric. It may not seem like progress to have
rewritten the metric in a form which is singular at $\tilde{U}=0$, but
this is an essential step in simplifying the task of studying
infalling strings.

We can rewrite this singular term in a more useful form by following
an argument given in \cite{hor:twsing}. Define
\begin{equation} \label{so}
\sigma_0 = \frac{1}{L} \int_0^L \sigma du, \quad G_0(u) = e^{\sigma_0 u},
\quad U_0(u) = -\frac{1}{2 \sigma_0 G_0^2},
\end{equation}
and
\begin{equation} \label{eta}
\eta_G = \frac{G}{G_0}, \qquad \eta_U = \frac{U}{U_0}.
\end{equation}
It follows that $\eta_G$ is a periodic function of $u$ with period
$L$, and therefore is bounded from above and below. This implies that
$\eta_U$ is also bounded from above and below. The divergent term can
be rewritten in terms of these functions,
\begin{equation} \label{oneU}
\frac{\dot{\sigma}}{R^2-1} = \frac{\dot{\sigma} G^2}{R^2 W^2} =
-\frac{\dot{\sigma} \eta_G^2 \eta_U}{2 \sigma_0 R^2 W^2} \frac{1}{U}
\approx -\frac{\dot{\sigma} \eta_G^2 \eta_U}{2 \sigma_0}
\frac{1}{\tilde{U}}.
\end{equation}
In the final step, we have taken $R\approx 1$ and $W$ approximately
constant. Thus, we see that this divergent term $\sim
1/\tilde{U}$. Let's write
\begin{equation} \label{delta}
\delta = -\frac{\dot{\sigma} \eta_G^2 \eta_U}{2 \sigma_0}.
\end{equation}
This coefficient is oscillating very rapidly near the horizon, as part
of it is periodic in $u$. At small values of $\tilde{U}$, one period
of $u$ occupies $ \Delta \tilde{U} < \tilde{U}$.

To simplify the analysis of the motion of test strings, we will
restrict to small inhomogeneities, i.e., $\delta$ small. This can be
motivated by considering a thermal ensemble of black strings with a
given total momentum. In such an ensemble, the typical state will only
have small inhomogeneities in the momentum distribution. More
precisely, $\dot{\sigma}_{rms} \sim \sigma_0^2/ r_0^2$
\cite{mar:seff}, which implies $\eta_G^2 \eta_U \sim 1$, and
$\delta_{rms} \sim \sigma_0/ r_0^2$. Thus, $\delta$ will be small if
we have chosen the background curvatures to be small.  

The metric simplifies to
\begin{equation} \label{appwavemet}
ds^2 \approx -d\tilde{U} d\tilde{V} + dX_\mu dX^\mu +
\frac{\delta}{\tilde{U}} (3 X_4^2 - X_i X^i)  d\tilde{U}^2.
\end{equation}
This metric describes a singular gravitational plane wave. Let us
briefly review the various approximations that have gone into this
result. Taking the three-sphere directions to be flat, and ignoring
the other finite contributions to the curvature, will be a good
approximation so long as our test objects are small compared to $r_0$
and $\sqrt{r_0/\sigma}$. Taking $q$ to be approximately constant is
valid sufficiently close to the horizon, and Marolf has argued
\cite{mar:seff} that when the inhomogeneities are small, it is valid
for the whole region of large curvatures.

We consider a string propagating in the metric (\ref{appwavemet}). We
quantize the string in light-cone gauge, $\tilde{U} = P \tau$, where
$\tau$ is worldsheet proper time. We will work in string units, so
$\alpha' =1$. If we decompose the transverse coordinates into modes,
\begin{equation} \label{modes}
X^\mu(\sigma, \tau) = \sum_n X^\mu_n(\tau) e^{in\sigma},
\end{equation}
the worldsheet field equations become
\begin{equation} \label{eq1}
\ddot{X}^i_n + n^2 X^i_n + \frac{\delta P^2}{\tilde{U}} X^i_n = 0 
\end{equation}
and
\begin{equation} \label{eq2}
\ddot{X}^4_n + n^2 X^4_n - \frac{3\delta P^2}{\tilde{U}} X^4_n = 0, 
\end{equation}
where an overdot now denotes $d/d\tau$. For $\tilde{U} \gg 0$ and
$\tilde{U} \ll 0$, the metric (\ref{appwavemet}) is approximately
flat, and the solutions reduce to combinations of the usual flat space
solutions $e^{\pm i n \tau}$.\footnote{Note that for $r_0 \gg1$, there
is a region inside the horizon where it is sensible to treat the exact
metric as approximately flat. I'm assuming that in the exact metric
(\ref{twmet2}), it is possible to propagate the string through the
singularity.} We will denote the region $U \ll 0$ by a subscript $<$
and the region $U \gg 0$ by a subscript $>$. We can define two
complete sets of solutions of (\ref{eq1},\ref{eq2}). The solutions
$u_{n<}, \tilde{u}_{n<}$ are defined to be pure positive and negative
frequency for $U \ll 0$, while $u_{n>}, \tilde{u}_{n>}$ are defined to
be pure positive and negative frequency for $U \gg 0$. There is a
linear relation between these two sets of states, which implies a
transformation between the initial and final mode creation and
annihilation operators, the Bogoliubov transformation
\begin{equation} \label{bog}
a^i_{n>} = A_n a^i_{n<} - B^*_n \tilde{a}^{i^\dagger}_{n<}, \qquad
\tilde{a}^i_{n>} = A_n \tilde{a}^i_{n<} - B^*_n a^{i^\dagger}_{n<}. 
\end{equation}
A string initially in the vacuum state will become excited on passing
through the wave. The excitation in a particular mode is given by 
\begin{equation} \label{Nn2}
\langle N_{n} \rangle = \langle 0_< | N_{n>} | 0_< \rangle = |B_n|^2. 
\end{equation}

To find the coefficients $B_n$, we propagate a positive-frequency
solution $u_{n<}$ through the plane wave and ask for its
negative-frequency part $\tilde{u}_{n>}$. The effective potential in
(\ref{eq1},\ref{eq2}) is far too complicated to allow us to solve this
problem exactly. However, if we consider large $n$, $\delta
P^2/\tilde{U} \ll n^2$ except for a very narrow region. Furthermore,
in this narrow region, $\delta$ is oscillating with a wavelength much
shorter than the wavelength of the mode. The area of a half-cycle of
this oscillation is given by 
\begin{equation} \label{area}
\int \frac{\dot{\sigma}}{R^2-1} d \tilde{U} = \int \frac{\dot{\sigma}
G^2}{R^2 W^2} d\tilde{U} = \int \frac{\dot{\sigma}}{R^2} du. 
\end{equation}
Thus, the areas of the half-cycles are roughly equal near the
horizon. We can therefore approximate the potential by a series of
closely-spaced delta functions of alternating sign. The propagation
through one such delta function gives a contribution to $B_n$ of order
$1/n$, but the contribution from two successive delta functions
cancels to leading order. Therefore, for the purpose of obtaining an
estimate of $B_n$ for large $n$, we can ignore this central part of
the potential and treat the rest by the Born approximation. This gives
\cite{hor:wave,vesa:sing}
\begin{equation} \label{born}
B_n \sim \frac{P}{2in} \int d\tilde{U} e^{-in\tilde{U}/P}
\frac{\delta}{\tilde{U}} \sim \frac{P \delta_{rms}}{2in} \int d z e^{-iz }
\frac{\delta}{\delta_{rms}z}. 
\end{equation}
In the second stage, we set $z = n\tilde{U}/P$. Because $\delta$ is
oscillatory in $u \sim \ln(\tilde{U})$, the form of $\delta$ as a
function of $z$ is essentially the same as the form as a function of
$\tilde{U}$. The last integral should therefore contribute just a
numerical factor, so $|B_n| \sim P \delta_{rms}/ n$. Thus, if a string
initially in its ground state falls through the singularity, each mode
will be excited to
\begin{equation} \label{Nn}
\langle N_{n} \rangle \sim \frac{\delta_{rms}^2P^2}{n^2}.
\end{equation}

Let us now calculate the total excitation of the string. There are
three important measures of this total excitation. The string mass is
given by
\begin{equation} \label{Ms}
\langle M_>^2 \rangle \sim \sum_{n=1}^{\infty} n \langle N_{n} \rangle,
\end{equation}
while the total number of modes is given by
\begin{equation} \label{N}
\langle N_> \rangle \sim  \sum_{n=1}^{\infty}  \langle N_{n} \rangle,
\end{equation}
and the average size of the string is given by \cite{mit:stat}
\begin{equation} \label{r}
\langle r_> \rangle \sim \sum_{n=1}^{\infty} \frac{1}{n}\langle N_{n}
\rangle.
\end{equation}
Thus, we see that while the total number of modes and the average size
of the resulting string state remain finite, the mass diverges once we
pass through the plane wave. Note that the divergence doesn't come
from a divergence in some individual $\langle N_{n} \rangle$, but
rather from the sum over arbitrarily high modes, each of which makes a
small contribution. The size of the individual $\langle N_{n} \rangle$
will depend on the size of the inhomogeneities, but the divergence
occurs so long as there is any non-zero inhomogeneity. Although the
analysis was only carried out for small inhomogeneities, making the
inhomogeneities larger is unlikely to soften the divergence. One might
worry that this divergence is just an artifact of the plane wave
approximation. That is, one might think that keeping subleading parts
of the curvature would lead to a different conclusion. However, this
does not seem likely, as such subleading terms describe curvature on
large scales, and the string remains small compared to those scales
even after it has passed through the singularity, so it should be
insensitive to the behavior on such scales.\footnote{However, the
authors of \cite{vesa:sing} argue that for plane waves which give
divergent answers, changes in the solution at arbitrarily large scales
will make the answers finite.} Thus, it seems that these solutions
really are singular from the perspective of test strings. 

\sect{Discussion}
\label{concl}

We have studied the singularity at the horizon of black strings with
traveling waves discovered in \cite{hor:twsing,kaloper:twsing}.  We
have seen that quantum corrections will be important at the would-be
horizon. By adding infalling matter to the solution, we obtain
invariants which diverge at the horizon. The presence of large
invariants is a fairly unambiguous sign that quantum corrections will
become important. However, it is not clear what the effect of these
quantum corrections will be. Since the classical solution predicts a
finite distortion for infalling test bodies, one might suspect that
their motion will not be much affected by the quantum corrections.

The quantum corrections {\em are} important for test objects in string
theory. It is possible to describe the propagation of test strings
using an approximate metric which has a plane-wave form. This
approximation is valid if the test strings cross the horizon with $q$
approximately constant (so we can ignore the $q$ dependence in the
metric), and they remain small (so we can ignore the curvature of the
three-sphere). The mass-squared of a string initially in its ground
state diverges as we propagate it through the singularity, due to
contributions from arbitrarily high modes.  Thus, it is not consistent
to use a classical picture of the background spacetime in the
neighbourhood of the singularity. One must replace the description
given here with some kind of quantum picture once the curvature
becomes sufficiently large. I have only treated the six-dimensional
black string with longitudinal wave. The extension to the
five-dimensional case is an essentially trivial exercise. I expect
similar behaviour would be seen with internal waves as well, although
it is not clear to me whether a plane-wave approximation is possible
in that case.

What do these results tell us about black hole physics? The typical
extreme string state will have some non-trivial momentum distribution,
and hence will form a black string with singular horizon. Here we have
found that we cannot evolve test strings beyond this singularity using
the classical solution, but must include quantum effects. The question
then becomes to what extent we can think of the appropriate quantum
description as representing a black string. 

The entropy of the classical black string solutions can be reproduced
in string theory at weak coupling \cite{hor:tw1}. This appears still
more mysterious in light of the present results. The presence of large
curvature invariants indicates that quantum corrections to the
geometry will be important, but the uncorrected geometry still gives
the correct density of states. It's unclear what this means, but it
may be suggesting that the full quantum solution is still in some
sense like a black string. Myers has argued for the alternative point
of view \cite{myers:grg}, that the presence of the singularity
indicates that this state is essentially unlike a black string. The
only way to obtain a black string at strong coupling is then to start
with a completely thermal ensemble at weak coupling, in which only the
total momentum is fixed. Thus, no information is lost in the
transition from weak to strong coupling.

A fuller understanding of this puzzle will likely be useful in
understanding the role of black hole and black string solutions in
quantum gravity. The presence of these singularities at least gives a
convincing argument that quantum effects will be important at the
horizon, which is surely necessary if such effects are to have any
relevance for the information loss problem. It should be borne in mind
that we have only discussed the extreme solutions. It is much more
difficult to discuss the effects of small changes in the metric on
non-extreme black holes, as such changes will dissipate over
time. Thus, the argument made here is not easy to extend to the
non-extreme case.

\bigskip
\bigskip
\centerline{\bf Acknowledgments}
\medskip

It is a pleasure to thank G.T. Horowitz, R.C. Myers and H. Yang for
discussions. This work was supported in part by NSF grant PHY95-07065,
and by the Natural Sciences and Engineering Research Council of
Canada.

\begingroup\raggedright\endgroup


\begin{thebibliography}{10}
\newcommand{\enquote}[1]{``#1''}

\bibitem{hor:stringc}
G.~T. Horowitz, \enquote{The origin of black hole entropy in string theory,}
  gr-qc/9604051.

\bibitem{mald:D-ent}
J.~M. Maldacena, {\em Black holes in string theory\/}, Ph.D. thesis, Princeton
  University (1996), hep-th/9607235.

\bibitem{strom:D-ent}
A.~Strominger and C.~Vafa, \enquote{Microscopic origin of the
  {B}ekenstein-{H}awking entropy,} Phys. Lett. {\bf B379}, 99 (1996),
  hep-th/9601029.

\bibitem{gar:cswave}
D.~Garfinkle and T.~Vachaspati, \enquote{Cosmic string traveling waves,} Phys.
  Rev. D {\bf 42}, 1960 (1990).

\bibitem{gar:tw}
D.~Garfinkle, \enquote{Black string traveling waves,} Phys. Rev. D {\bf 46},
  4286 (1992), gr-qc/9209002.

\bibitem{hor:tw1}
G.~T. Horowitz and D.~Marolf, \enquote{Counting states of black strings with
  traveling waves,} Phys. Rev. D {\bf 55}, 835 (1997), hep-th/9605224.

\bibitem{hor:tw2}
G.~T. Horowitz and D.~Marolf, \enquote{Counting states of black strings with
  traveling waves. 2,} Phys. Rev. D {\bf 55}, 846 (1997), hep-th/9606113.

\bibitem{hor:twsing}
G.~T. Horowitz and H.~Yang, \enquote{Black strings and classical hair,} Phys.
  Rev. D {\bf 55}, 7618 (1997), hep-th/9701077.

\bibitem{kaloper:twsing}
N.~Kaloper, R.~C. Myers, and H.~Roussel, \enquote{Wavy strings: Black or
  bright?} Phys. Rev. D {\bf 55}, 7625 (1997), hep-th/9612248.

\bibitem{mar:seff}
D.~Marolf, \enquote{Statistical effects and the black hole / d-brane
  correspondence,} hep-th/9705063.

\bibitem{hor:prop}
G.~T. Horowitz and S.~F. Ross, \enquote{Properties of naked black holes,}
  hep-th/9709050.

\bibitem{hor:wave}
G.~T. Horowitz and A.~R. Steif, \enquote{Strings in strong gravitational
  fields,} Phys. Rev. D {\bf 42}, 1950 (1990).

\bibitem{gib:wave}
G.~W. Gibbons, \enquote{Quantized fields propagating in plane-wave spacetimes,}
  Commun. Math. Phys. {\bf 45}, 191 (1975).

\bibitem{vesa:sing}
H.~J. de~Vega and N.~Sanchez, \enquote{Strings falling into space-time
  singularities,} Phys. Rev. D {\bf 45}, 2783 (1992).

\bibitem{mit:stat}
D.~Mitchell and N.~Turok, \enquote{Statistical properties of cosmic strings,}
  Nucl. Phys. {\bf B294}, 1138 (1987).

\bibitem{myers:grg}
R.~C. Myers, \enquote{Pure states don't wear black,} gr-qc/9705065.

\end{thebibliography}
\end{document}